\documentclass[useAMS,usenatbib]{mn2e}

\usepackage{graphicx}
\usepackage{natbib}

\title[Instabilities in accretion discs]{On different types of instabilities in black hole accretion discs. Implications for X-ray binaries and AGN}
\author[A. Janiuk \& B. Czerny]{Agnieszka Janiuk$^{1}$\thanks{E-mail: agnes@cft.edu.pl}, Bo\.zena Czerny$^{2}$\\
$^{1}$ Center for Theoretical Physics, Al. Lotnikow 32/46,
02-680 Warsaw,Poland\\
$^{2}$N. Copernicus Astronomical Center, Bartycka 18, 00-716
Warsaw, Poland\\}

\begin{document}

\date{Accepted  Received in original form }

\pagerange{\pageref{firstpage}--\pageref{lastpage}} \pubyear{2010}

\maketitle

\label{firstpage}

\begin{abstract}

We discuss two important instability mechanisms that may lead to the 
limit-cycle oscillations of
the luminosity of the accretion disks around compact objects: ionization 
instability and
radiation-pressure instability. Ionization instability is well established 
as a mechanism of X-ray novae
eruptions in black hole binary systems but its applicability to AGN is still 
problematic. 
Radiation pressure theory has still very 
weak observational background in any of these sources. In the present paper 
we attempt to confront
the parameter space of these instabilities with the observational data. 
At the basis of this simple survey
of sources properties we argue that the radiation
pressure instability is likely to be present in several Galactic sources 
with the Eddington ratios above 0.15,
and in AGN with the Eddington ratio above 0.025. Our results favor the 
parameterization of the viscosity
through the geometrical mean of the radiation and gas pressure both in 
Galactic sources and AGN. More
examples of the quasi-regular outbursts in the timescales of 100 seconds 
in Galactic sources, and hundreds of
years in AGN are needed to formulate firm conclusions. We also show that 
the disk sizes in the X-ray novae 
are consistent with the ionization instability. This instability may 
also considerably
influence the lifetime cycle and overall complexity in
the supermassive black hole environment. 

\end{abstract}


\section{Introduction}
\label{sec:intro}

The description of the viscous torque through
the $\alpha$ parameter by \cite{ss73} boosted the modeling of the disk 
accretion onto the central object, with application in various fields, 
from young stellar objects through stellar close binary systems to active 
galactic nuclei. The parameterization was justified, but not based on a 
specific well developed theory. However, the properties of the stationary
disk flow weakly depend on this assumption. The spectra models were
totally unaffected by the viscosity unless either departure from the Keplerian 
motion or departure of the emission from a local black body 
were taken into account. 

However, the stability of disk models does depend significantly on the 
adopted description of the viscosity. First, it was noticed that the 
assumption of the
proportionality of the viscous troque to the total  (i.e. gas plus radiation) 
pressure leads to the viscous \cite{pringle74} and thermal \cite{lightman74} 
instabilities. 
Later,
the instability of the outer gas-pressure dominated parts of the disk 
was discovered, in the region of partially ionized hydrogen and helium
(\cite{meyer81}, \cite{smak84}). 
Such instabilities should lead to semi-regular periodic
outbursts. Therefore, observations of objects containing accretion disks
can be used to test the assumptions about the viscosity law.

Such confrontation of the models and theory is being done in case of the 
ionization instability. The early model development was actually motivated by 
the need to explain the dwarf novae outbursts in cataclysmic variables. 
This instability is also a leading explanation of the X-ray novae outbursts
(\cite{can82}; for a review see \cite{Lasota01}).
It may also apply to active galaxies (\cite{lin86}, \cite{mineshige90})
although observational tests are at the early stage of development.

The presence of the radiation pressure instability is not proved yet 
(see e.g. \cite{done07}). The
understanding of the true nature of the viscosity as being due to the
magnetorotational instability \cite{balb91} did not simply set the issue. 
The limit cycle cannot be seen in 3-D simulations of this instability since 
the authors cannot follow the global evolution
of the disc in a viscous timescale and the radial propagation of heating and
cooling fronts is here neglected. However, recent computations indicate 
that radiation pressure contributes to the viscous torque and the
instability may be there \cite{hirose09b}. The comparison of the
predictions of this instability with observational data was done only in a few
papers so far. 

In this work we propose to study further the models of 
the accretion disc instabilities in a global picture, as well as to better 
support them observationally. In case of the radiation pressure 
instability, we use two models: one is 
based on the assumption of the viscoscous 
torque proportional to the total pressure and the other is based on the 
assumption of the viscosous torque proportional to geometrical mean of the 
gas and radiation pressure. We also include the cooling term due to the 
outflow. We mark the instability strips in the disk radius - acccretion rate
plane and we compare
them with the observed properties of the X-ray lightcurves of accreting black 
holes in binary systems, taken from the literature. 
We claim that the observational constraints
for the thermal disc instability and the limit-cycle behaviour is much more than
the one famous example of the microquasar GRS 1915+105. We give some examples 
of other sources and discuss further need for the detailed observational 
studies for both Galactic individual X-ray sources and AGNs.

This article is organized as follows. In Section \ref{sec:results} 
we discuss the theoretical background for the radiation pressure
and partial hydrogen ionization instabilities.
We also present the results for the extension and overlapping
of the unstable regions in the accretion discs. In particular, we
focus on the constraints for the accretion rate and jet efficiency 
which would be adequate for the astrophysical black hole discs to become 
unstable
for one or both types of instabilities. 
These results are based on
the numerical codes developed by ourselves and discussed in detail
in a series of previous works.
In section \ref{sec:obs} we present the observational constraints for the
disc instabilities found for a number of Galactic black hole binaries. 
We also discuss the supermassive black hole AGNs and some observational 
constraints found in the literature.
In Section \ref{sec:diss} we give a summary and conclusions.

\section{Disc instabilities}
\label{sec:results}

\subsection{Radiation pressure}
\label{sec:prad}

The black hole accretion disc with the
classical heating term proportional to the pressure with 
the $\alpha$ coefficient (\cite{ss73}) is
subject to the thermal and viscous instability when the radiation
pressure dominates over the gas pressure. 
This occurs in the innermost radii of the accretion disc around a
compact object (in case of a white dwarf such a region cannot be present).
Radiation pressure instability of classical alpha models of 
\cite{ss73} was noticed very early (\cite{lightman74},
\cite{pringle74}) 
and it was fully analyzed by \cite{ss76}.

The time evolution of the system and its stability is
governed by the accretion rate outside the unstable region
(i.e., the mean accretion rate). If the accretion rate is low,
then the disc remains cold and stable, with a constant low luminosity.
If the accretion rate is large, then the whole disc becomes hot and is
stabilized by advection, i.e. enters a slim disc solution (\cite{abr88}).
However, for the intermittent accretion rates, larger than some critical value,
the unstable mode activates.
In this case the source enters a cycle of bright, hot states,
separated by the cold, low luminosity states. 
The outburst amplitudes and durations
are sensitive to black hole mass, viscosity parameter and 
the mean accretion rate.  
Also, the heating prescription is important here. 
If the viscous heating is proportional to the total pressure, the outburst
amplitudes are very large, however, they can be reduced if the heating is 
proportional to the square root of the gas times the total pressure.
If we assume the heating proportional only to the gas pressure,
the instability disappears. In case of the geometrical mean of the two 
components the parameter space of the instability is greatly reduced.

Preliminary shearing-box 3D simulations replacing the alpha
viscosity with a physical (magnetic) viscosity mechanism indicated
that there is no thermal runaway even when the radiation pressure 
is 10 times larger than the gas pressure (\cite{hirose09a}). 
However, the same authors (\cite{hirose09b}) in their follow-up
work already suggested the possibility of radiation pressure instability 
in some of their calculations, as the unstable solutions may be 
seen on the surface density - effective temperature plot. 
The limit cycle cannot be seen in those simulations since they do not follow the global evolution of the disc in a viscous timescale and the radial 
propagation of heating and 
cooling fronts is here neglected. Full time-dependent computations of the 
global evolution can be only performed with a simple viscosity 
parameterization, and in such computations a limit-cycle behaviour is seen, 
with disc alternating between the hot and cold states (e.g. \cite{nayak00}, 
\cite{janiuk02}, \cite{janiuk07}, \cite{czerny09}). 

Observationally, the situation is far from clear (see e.g. the review by 
\cite{done07}). Several authors suggested that the radiation pressure instability is  an attractive explanation of the regular outbursts lasting a 
few hundreds of seconds observed in the microquasar GRS 1915+105
(e.g. \cite{taam97}, \cite{deegan09}). The radiation pressure instability is
the only model which explains the absence of the direct transitions from the 
state C to the state B in this source.  The lightcurves of 
some other objects also show the fluctuations in a form of a limit cycle
on the appropriate timescales. An interesting example is X-ray pulsar
GRO J1744-28 (\cite{can96}, \cite{can97}) with periods of very high accretion 
rate and low magnetic field which allows for the presence of the inner 
radiation-pressure dominated part of the disk. This instability operates when 
the radiation pressure is important, so it is expected only in high 
Eddington ratio Galactic sources and AGN. However, for many sources accreting 
at very high rates the limit cycle
oscillations have not been reported. Comparison of the Eddington ratios of
the stable sources and those showing fast (100 - 1000 s) regular outbursts is 
a key test of the correct viscosity parameterization.

\subsection{Partial Hydrogen ionization}
\label{sec:ioniz}
Another type of the 
thermal - viscous instability occurs due to the partial
ionization of hydrogen and helium. The unstable zone is typically present in 
the outer part of the disc, where the effective temperature is of order of 
6000 K. The ionization instability was first found by \cite{meyer81} 
and \cite{smak84} in the disks surrounding white dwarfs and it is at present 
the accepted explanation of the dwarf novae
and X-ray novae outbursts.
The numerical models presented in a number of works 
(e.g., \cite{dubus01},\cite{janiuk04}; for a review, see \cite{Lasota01})
studied the one or two (1+1) dimensional models 
of the global evolution of accretion discs under the thermal and viscous 
instability and confirmed the possibility of the limit cycle behaviour.

 The hydrogen ionization instability is thus an example of a firm agreement
between the observations and theory. A recent example is SU UMa-type 
dwarf nova V344 Lyr observed by Kepler satellite, for which its 
outbursts and superoutbursts have been modeled with this instability 
(\cite{can10}).

This instability also likely applies to active galaxies 
(\cite{lin86}, \cite{mineshige90}) although this is still uncertain.
The models of the outbursts were further developed by  \cite{siem96}, 
\cite{janiuk04}, and they were applied to statistics of AGN by \cite{sie97}.  
\cite{MenQ01} and \cite{hameury09} argued that the amplitude of such outbursts will be small but the evaporation of the inner disk enhances the amplitude considerably \cite{janiuk04} and prolongs the qiescent state \cite{hameury09}.   

In this case the situation is much similar to the above, 
as the disc cycles between two states. The
hot and mostly ionized state of a large local accretion 
rate is intermittent with a
cold, neutral state of a small local accretion rate. 
Again, the quantitative outcome of the model is governed by the assumed
external (mean) accretion rate, viscosity and mass of the central object.

\subsection{Location of the unstable zones}
\label{sec:locations}

We calculated the steady-state models of the accretion disc structure, 
for two exemplary values of the black hole mass, characteristic for
Galactic sources (10 $M_{\odot}$) and AGN ($10^{8} M_{\odot}$).
The models are based on the vertically averaged equations for the 
energy balance between the viscous heating and radiative as well as 
advective cooling and hydrostatic equilibrium. The heating term, governed by
the viscosity parameter $\alpha$, is in the radiation pressure
dominated region assumed proportional to either
the total pressure, or to the square root of the total times the
 gas pressure.
In the gas pressure dominated region, located at larger distances,
the heating is assumed proportional 
only to the gas pressure. We calculate here
the vertical profiles of temperature, density and pressure,
using the opacity tables that cover the temperature range relevant for partial
hydrogen ionization, including the presence of dust and molecules
 (see details in \cite{roz99}).

The basic parameter of each stationary
 model is the global (external) accretion rate, 
through which we determine the total energy flux dissipated in the 
disc at every radius $r$. 
Once the effective temperature and surface density are determined at 
every disc radius, we find the stable solutions, i.e. the accretion rates 
for which the slope of $T-\Sigma$ (or $\dot M -\Sigma$) 
relation is positive, and the unstable 
solutions, with the negative slopes. 
In other words, the ``S-curve'' is plotted locally at a number of disk 
radii, and we search for the critical $\dot m$ points at which the 
curve is bending. These points limit the maximum and minimum 
values of accretion rates for which at a given radius the disk will be 
unstable.
In turn, we determine the range of radii, for which at a given
global accretion rate the disc is unstable first due to the
radiation pressure and then to the ionization instability.

\begin{figure}
\includegraphics[width=8cm,height=8cm]{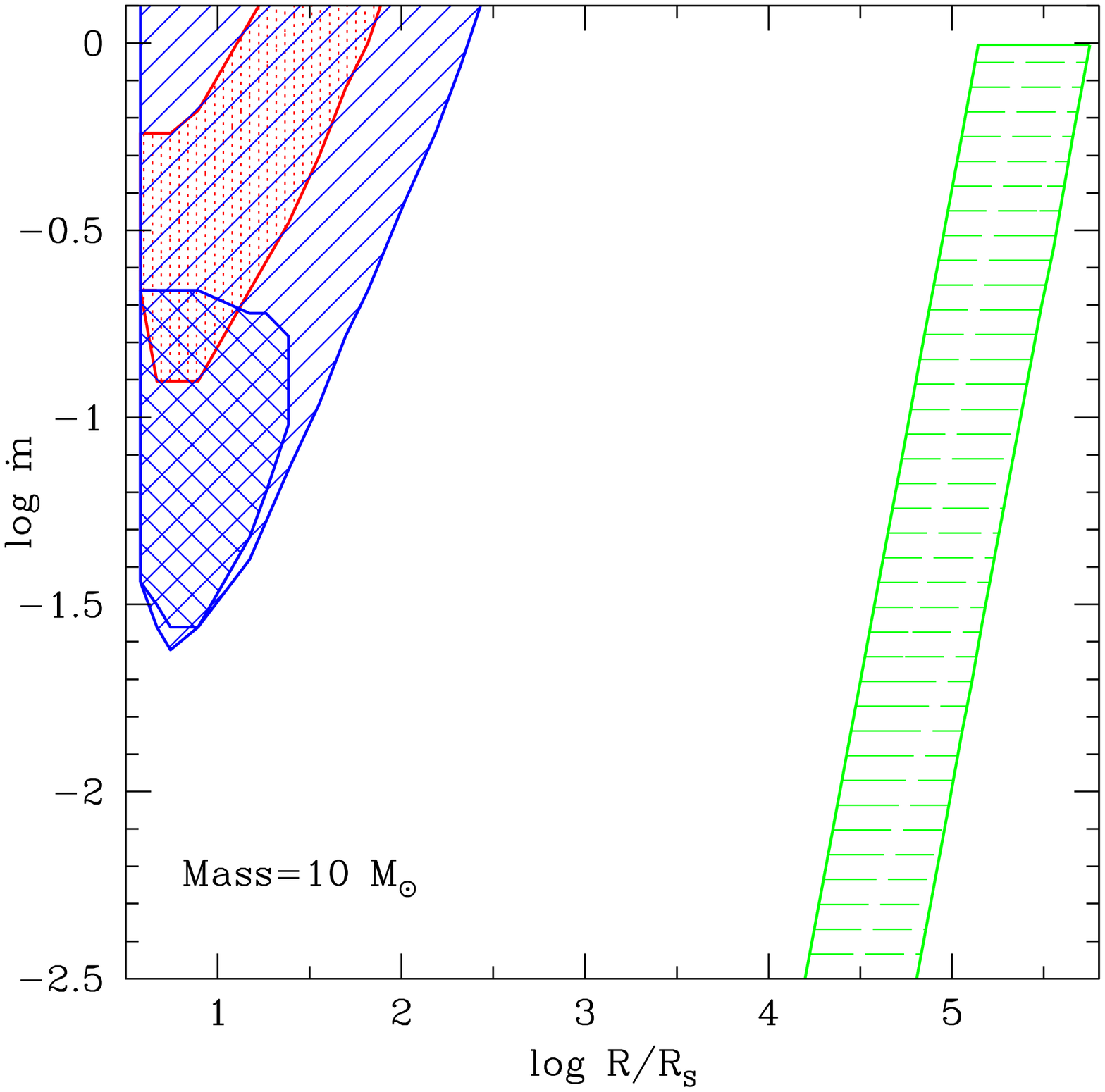}
\includegraphics[width=8cm,height=8cm]{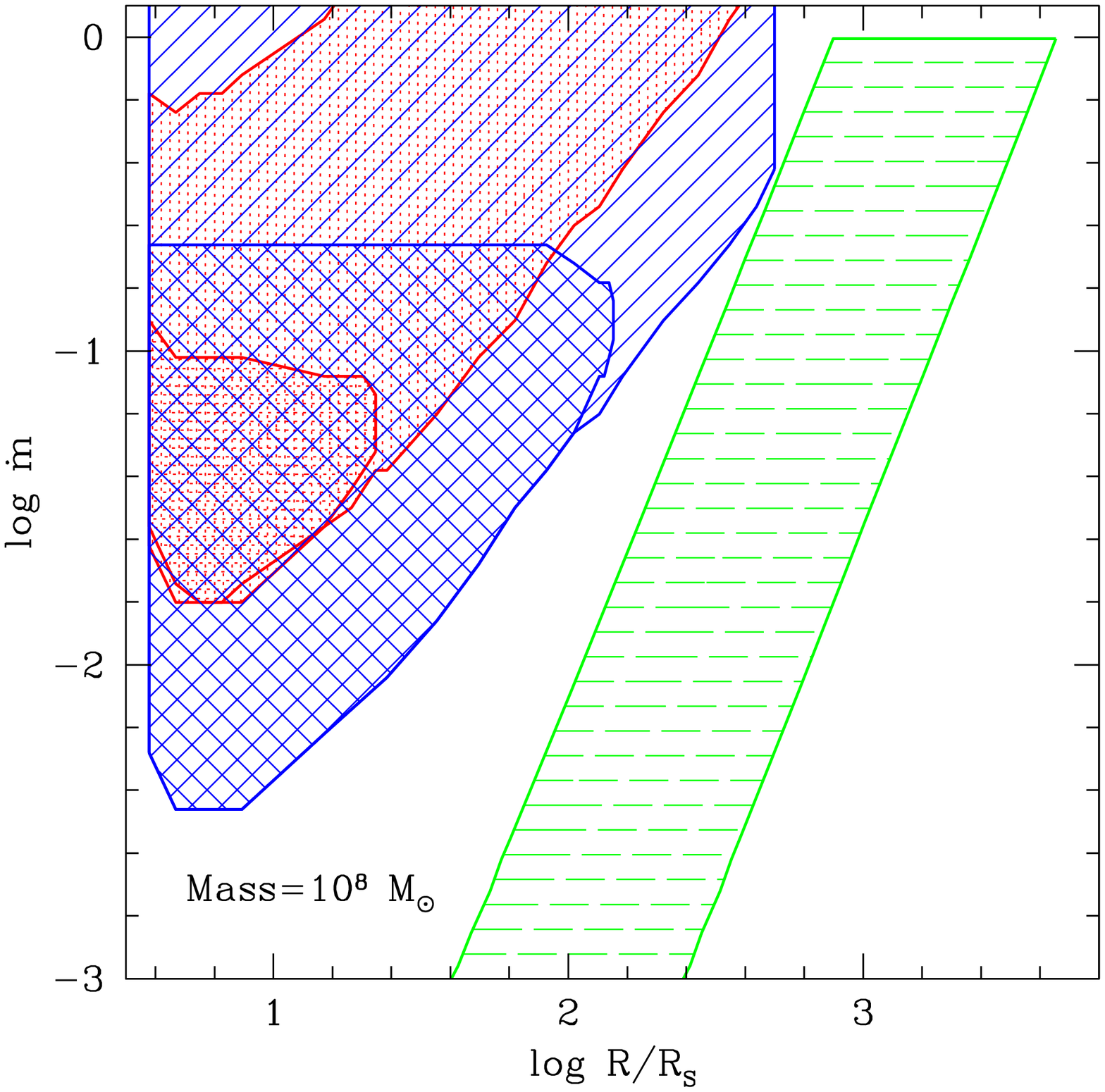}
\caption{The extension of the radiation pressure (solid and dotted lines)
and hydrogen ionization (green; dashed lines) unstable zones, depending on mean 
accretion
rates (Eddington units). The results are for two heating prescriptions:
$\alpha P_{\rm tot}$ (blue; solid lines) and $\alpha \sqrt{P_{\rm gas}P_{\rm tot}}$
(red; dotted lines).
The crossed regions mark the results for a
 non-zero fraction of jet power, described by Eq. (\ref{eq:jet})
with $A=25$.
The black hole mass is
$M = 1\times 10^{8} M_{\odot}$ (bottom) and $M = 10 M_{\odot}$ (top). 
The viscosity is $\alpha=0.01$. } 
\label{fig:topo}
\end{figure}

For the latter, the unstable strip is located at the outskirts of
the disk. Obviously, if the instability arises at the outer disk, the front 
will then propagate inwards to much smaller radii. However, if the
disk size is smaller than the inner edge of the unstable strip plotted
in Figure \ref{fig:topo}, no ionization instability outbursts should take 
place. 
Our results, based on the detailed vertical structure calculations, are
consistent with the simplified formulae given in the
Appendix  of \cite{Lasota01}, with respect to the inner boundary of the 
ionisation instability strip.
The outer edge we determined is somewhat larger, due to a different
opacity tables in our model which include absorption on molecules, e.g. 
molecular hydrogen, as described in \cite{roz99}.

In Figure \ref{fig:topo} we show the maps of the disc instabilities
for the two chosen black hole masses, on the plane 
radius vs. global accretion rate (in dimensionless units).
In addition, we distinguish the two possible stabilizing mechanisms for
the radiation pressure instability: the heating prescription and 
the possibility of energy outflow to the jet. The latter is parameterized by
the following function:
\begin{equation}
\eta_{\rm jet} = 1 - {1 \over 1 + A \dot{m}^{2}}
\label{eq:jet} 
\end{equation}
and the jet outflow acts as a source of additional cooling 
(\cite{nayak00}, \cite{janiuk02}).

The jet outflow reduces the size of the unstable zone
for large accretion rates, as well as limits the instability to operate below
some threshold maximum $\dot m$. This rate is of course sensitive to our adopted parameter
for the jet strength. The Figure \ref{fig:topo} shows the case of
a very strong jet, with $A=25$ and in this case the limiting accretion rate is
about 20\% of the Eddington rate. For a 10 times weaker jet the limiting 
accretion rate is about 3 times larger.

The viscosity parameter, $\alpha$, only moderately affects the results for 
the radiation pressure instability, apart from the timescales of the limit cycle. 
The Figure \ref{fig:topo} presents the conservative 
case of a very small viscosity, $\alpha=0.01$, for which the extension of the 
unstable
zone is small. If the viscosity is larger, the unstable zone size 
is also increased, for instance 
for $\alpha=0.1$ the outer radius of the instability is larger by a factor 
of $\sim 1.25$.
This also means that the minimum accretion rate for which the disc
is unstable, is slightly smaller. However, in this smallest accretion rates
the unstable zone is very narrow.
For a very narrow width of the zone, the instability does not work, because the
front does not propagate,
and we have only marginally stable solutions (\cite{szusz97}).
For instance, \cite{hameury09} find that the heating and cooling fronts
of the ionization instability 
do not propagate strongly enough in their model to account for the large 
luminosity oscillations in AGN.

The outburst cycles caused by the ionization instability are very sensitive
to the viscosity parameter. As was shown already for dwarf novae,
the amplitudes consistent with observations can be obtained only in
the models with non-constant viscosity, i.e. $\alpha$ in the hot state must be
 larger
than in the cold state. This may also be the case in the X-ray binaries, 
however in AGN the situation may be different (see \cite{janiuk04}).

Thus the fast radiation pressure instability is expected to operate for
an average accretion rate  higher than a certain lower limit, 
mostly dependent on
the adopted viscous scaling. There is an upper limit as well
 if the outflow is strongly 
increasing function of the Eddington ratio. 
The ionization instability should operate if
the disc is large enough to show a partially ionized hydrogen zone for a 
given average accretion rate. Confronting the observational constraints 
with model predictions will in turn allow us to 
find the constraints for the viscosity parameterization and the role of the 
outflow. In the
next sections we make a preliminary step in this direction.

\section{Observational constraints}
\label{sec:obs}

The Galactic X-ray binary systems are variable
in wide range of timescales. First, the transient X-ray sources
undergo their X-ray active states on timescales of years.
Second, some sources exhibit X-ray periodic variability  
on timescales of months.
Third, some of the most luminous sources are variable
in timescales of tens-thousands of seconds. Finally, many of
X-ray binaries undergo quasi-periodic oscillations. Direct comparison of
these data and the models is not simple: even the identification of the
type of the variability with the mechanism is not unique.

In Table \ref{tab:binaries} we summarize the properties 
of the exemplary, best studied X-ray binary sources found in the
literature, which in our opinion may display radiation pressure or 
ionization instability. 
We list their characteristic variability timescales and amplitudes,
estimated Eddington ratios and disc sizes, as well as we indicate
a possible instability mechanism responsible for the variability, whenever
it is in agreement with our computations presented in Sec. \ref{sec:results}.
The maximum disc radius is estimated based on the 60\% of the Roche lobe
size, from the simplified formula given by \cite{pac71}, 
whenever we had the data for the system 
orbital parameters.

\begin{table*}
\caption{Sample of the black hole X-ray binary sources. $\Delta T$ is the estimated duration of an outburst,
and $ F_{max}/F_{min}$ is its amplitude. $R_{d}/R_{s}$ is the estimated disc size in Schwarzschild units.
The observations were found in the literature and taken from http://xte.mit.edu/}
\begin{tabular}{lcccccr}
\hline
\hline
Source          & $\Delta T$  & $ F_{max}/F_{min}$ & $\dot M/\dot M_{Edd}$ & $R_{d}/R_{s}$   & Instability & Ref. \\
\hline
\hline
A0620-00                & 150 days     &    300          &$10^{-2} - 3$ &$4.8\times 10^{5}$ & Ioniz.   & 14 \\
GRS 1915+105            & 20-100 yrs   &    $> 100$      & 0.25-0.7 & $6.3\times 10^{5}$  & Ioniz.   & 2 \\
GRS 1915+105            & 100-2000 s   &    3-20         & as above & as above           & $P_{rad}$ & 1,3,19\\
GS 1354-64              & $\sim$ 30 d  &    $> 20 $      & 0.1-1.8  & $1.8\times 10^{5}$  & Ioniz.    &  33 \\
GS 1354-64              & $\sim$ 20 s  &    1.5-2        & as above &  as above         & $P_{rad}$ & 4,20,21, 35\\
XTE J1550-564           & 200  d       &    300          & $\sim 0.15$ & $1.3\times 10^{5}$ & Ioniz & 36 \\
XTE J1550-564           & $\sim$2000 s &    1.5          & $\sim 0.15$ & as above & $P_{rad}$ & 5, 22 \\
GX 339-4                & 100-400 days &   75            & $< 0.05$ & $1.6\times 10^{5}$ & Ioniz.& 6, 23, 24\\
GRO J0422+32            & 200 days     &   $>30$         & $0.002 - 0.02$& $4.8\times 10^{4}$  & Ioniz.& 7, 25 \\
GRO J1655-40            & 20-100 days  &   16  & $5\times 10^{-4}-0.45$& $1.0 \times 10^{5}$& Ioniz.& 8, 26, 27 \\
GRO J1655-40            & 0.1-1000 s   &   7.5           & as above   & as above           & $P_{rad}$  & 8, 32 \\
4U 1543-47              &  50 days     &   300 & $4.5\times 10^{-4}-0.04$ & $9.6\times 10^{4}$ & Ioniz.&  9 \\
GS 1124-684             & 200 days     &   24     &  $\sim 10^{-4}$ - $\sim 1.0$
                            & $5.2 \times 10^{4}$ & Ioniz.    & 6,14 \\
GS 2023+338             & 150 days     &    $>100$       & $0.01- 1.0$&  $3.8 \times 10^{5}$& Ioniz.&10, 29, 30 \\
GS 2023+338             & 60 s ?       &   500           & as above &  as above            &  $P_{rad}$ & 10 \\
SWIFT J1753.5-0127      & 150 days     &   10            &  0.03   &  $2.0 \times 10^{4}$ & Ioniz.    & 17, 31 \\
4U 1630-472             & 50-300 days  &   60            &            &                      & Ioniz.  & 28 \\
GRS 1730-312            & 6 days       &   200           &            &                      & Ioniz.  &  11\\
H 1743-322              & 60-200 days  &   100           &            &                      & Ioniz.  &  12\\
GS 2000+251             & 200 days     &   240           &            &                      & Ioniz.  &  6\\
MAXI J1659-152          & 20 days      &   15            &            &                      & Ioniz.  &  \\
CXOM31 J004253.1+411422 &$ >30$ days   &   $>300$          &          &                      & Ioniz.  &  13\\
XTE J1818-245           & 100 days     &   40            &            &                      & Ioniz.    & 15 \\
XTE J1650-500           & 80 days      &   120           &            &                      & Ioniz.    & 16\\
XTE J1650-500           & 100 s        &   24            &            &                      & $P_{rad}$  & 18\\
\hline
\end{tabular}
\\
\small{ $^{1}$ Wu et al. 2010; $^{2}$Deegan et al. 2009; $^{3}$Taam et al. 1997;
$^{4}$ Revnivtsev et al. 2000;$^{5}$ Homan et al. 2001; $^{6}$ Tanaka \& Shibazaki 1996; 
$^{7}$ van der Hooft et al. 1999; $^{8}$ Harmon et al. 1995; 
$^{9}$ Gliozzi et al. 2010; $^{10}$ in't Zand et al. 1992; $^{11}$ Trudolyubov et al. 1996; 
$^{12}$ Motta et al. 2010; $^{13}$ Garcia et al. 2010; $^{14}$ Esin et al. (2000); $^{15}$
Cadolle Bel et al. 2009; $^{16}$ Corbel et al. 2004,$^{17}$ Soleri et al. 2008; 
$^{18}$  Tomsick et al. 2003; $^{19}$ Belloni et al. 2000; $^{20}$ Kitamoto et al. 1990; $^{21}$ 
Casares et al. 2004; $^{22}$ Sobczak et al. 2000; $^{23}$ Hynes et al. 2003; $^{24}$ Miller et al. 2004;
$^{25}$ Shrader et l. (1997); $^{26}$ van Paradijs 1996; $^{27}$ Kolb et al. 1997; $^{28}$ Buxton \& Bailyn 2004
$^{29}$ \.Zycki et al. 1997; $^{30}$ \.Zycki et al. 1999; $^{31}$ Zhang et al. 2007; $^{32}$ Greiner 1994; 
$^{33}$ Brocksopp et al. 2001; $^{34}$ Osterbroek et al. 1997; $^{35}$ Cui et al. 1999; $^{36}$ Sobczak et al. 1999 
 }
\label{tab:binaries}
\end{table*}

The estimates given in Table \ref{tab:binaries} should be treated only as
indications. The information comes mostly from the literature and frequently
relies on quantitative description. All objects in Table  \ref{tab:binaries}
display ionization instability since we selected objects classified as X-ray novae
and there is very little uncertainty in the establishment of their nature. 
Most of them show large amplitude outbursts lasting days. The ratio of the maximum 
to minimum flux was estimated from the peak and the emission level 
at the end of 
the outburst so it represents the lower limit - only some of the X-ray 
novae have 
clear detection in the quiescence. Clearly, more careful observational 
analysis is 
needed to better study the amplitude pattern during the ionization 
instabilities in these sources.

However, from the comparison of the accretion rates and disc sizes
with our map presented in Figure \ref{fig:topo}
one can already infer some information about the individual sources.
All but one among the sources are well within the instability strip, 
so the instability operates as expected. Only one source, 
SWIFT J1753.5-0127, is at the inner border
of the instability strip. However our disc size estimation is based
on the black hole mass. If this mass is larger than current value inferred from the 
mass function, the unstable strip will be broader in its accretion disc.
The amplitude of the outbursts in this source is rather small compared to the
other X-ray novae in Table \ref{tab:binaries} which would be consistent
with the narrow instability strip
 due to too small disk radius. Further studies of
this exceptional source can provide key tests of the exact location of the 
instability zone. 

We also indicated in Table \ref{tab:binaries} which sources are promissing
candidates for the presence of the radiation pressure instability. In selecting
them we paid attention to the possible detection of exceptionally low frequency QPO
or just reports of outbursts or variability in timescales longer than 10 s.
Any faster variability than 10 s in unlikely related to radiation pressure instability
and for such fast QPO there are other mechanisms under consideration.

GRS 1915+105 is the most obvious candidate, with its semi-regular outbursts 
in timescales of 100 - 2000 s present in several among the brighter characteristic 
states (\cite{belloni00}). Those outbursts were already modeled 
by several authors as caused by the radiation pressure instability.
 
\begin{figure}
\includegraphics[width=8cm,height=8cm]{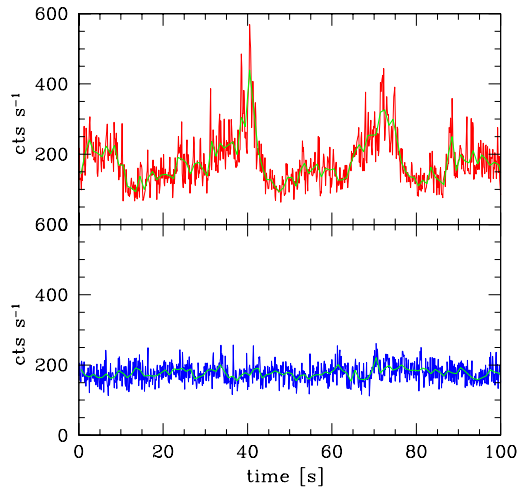}
\caption{Top panel: an RXTE lightcurve of GS 1354-64, observed on 05/12/1997.
Bottom panel: a lightcurve of Cyg X-1, observed by RXTE in the hard state,
rescaled to the mean countrate of GS 1354-64. The
time bin is 0.1 s in both data.}
\label{fig:compar}
\end{figure}

However, fast outbursts are apparently present in several other sources. We show
a specific example.
In Figure \ref{fig:compar} we show an exemplary lightcurve 
of GS 1354-64, which we extracted from RXTE data archive. Clearly,
a periodicity of the $\sim 20$ s outbursts is visible in the data.
The profiles with a slow rise and fast decay are characteristic for the
limit cycle oscillations in the radiation pressure instability.
For comparison, the lightcurve of Cyg X-1 in its hard state 
is plotted in the 
bottom panel of the Figure \ref{fig:compar}.
We see here a very stable X-ray emission and 
no signatures of the cyclic outbursts. Other sources were 
selected at the basis of their description in the 
literature. The selction is thus not completely objective or uniform
but may serve as a guide of the viability of the approach.

Having devided 
the Galactic sources into those which possibly show radiation
pressure instability and those which seem
 stable we can compare the Eddington ratios
within the two groups.

Sources with the Eddington ratio below 0.03 are stable. Examples are GRO J0422+32, GX 339-4,
as well as Cyg X-1. Among the unstable sources,
the object XTE J1550-564 has the smallest Eddington ratio, 0.15. The instability
seen in this source, however, is likely marginal. \cite{cui99} reported 82 mHz oscillations,
whith the frequency later increasing to a few Hz. Low frequency oscillations 
in this source were recently studied by \cite{rao10}, and they report periods varying between 2 and 10 Hz which is far too high for the radiation 
pressure instability. The interesting transition hinting for an instability was reported
by \cite{homan01}. In the MJD 51,254 observation,
 when the source was still very bright,
the luminosity suddenly increased without a change in the color. 
Whether indeed this
single transition hints for the radiation pressure instability or not,
 the source
likley defines the lower limit for the radiation pressure instability to operate.

In Table \ref{tab:binaries} we do not see sources which have very large 
Eddington ratio
and are stable against the radiation pressure instability. As we mentioned 
above, 
GRS 1915+105 is a good example of showing outbursts even at Eddington ratio 
close to 1
so it seems we have no upper limit for the Eddington ratio in the case of 
radiation pressure
instability.

Thus, observationally, the radiation pressure instability should operate between 
the Eddington ratio 0.15 up to 1 or more. Comparing this with the several theoretical 
possibilities plotted in Fig. \ref{fig:topo} we can draw certain conclusions.

First, only the viscosity prescription $\alpha \sqrt{P_{\rm gas}P_{\rm tot}}$ is consistent
with the lower limit for the radiation pressure instability, as the unstable region then 
extends from the Eddington 0.16 up. The prescription  $\alpha P_{\rm tot}$  would allow
instability to operate at too low luminosity.

Second, too efficient cooling by the jet is also ruled out. 
The cooling operates similarly in both cases of viscosity parameterization
and stabilize the disc. For the adopted values of the jet efficiency parameter,
the disc is stable for 
Eddington ratio above 0.22, which is clrearly inconsistent with observations.
Therefore, the parameter $A$ is Eq.\ref{eq:jet} of the disk-jet coupling 
must be significantly lower 
than this exemplary value of $A=25$. However the jet is by no means
 excluded and still can
carry a substantial energy, because in the case of
equipartition between the disk and jet radiation, for the 
Eddington accretion rate $\dot m =1$ the jet coupling constant 
equal to $A=1$ would be enough. 
 
The parameterization $\alpha \sqrt{P_{\rm gas}P_{\rm tot}}$  has an additional advantage of 
reducing the outburst amplitude in comparison to $\alpha P_{\rm tot}$. 
Most of the candidate sources for radiation pressure instability show rather 
low to moderate amplitudes, from factor 2 to 20. 
Only one source - GS 2023+338 - shows huge outbursts, with the factor of 
500 brightenings in timescales of 60 seconds.  \cite{zand92} interpreted 
this short timescale variability as caused by variable absorption. The 
behaviour of this source is exceptional and puzzling.

When studying the instabilities in the supermassive black hole environment, we
usually cannot directly observe a duty cycle of a one single object,
since the black hole masses are large and the expected timescales are very long.
Instead, the statistical studies are useful here and we can
find an evidence for the source episodic activity (e.g. \cite{czerny09}).
However, the exceptional object is NGC 4395, with the black hole mass
of $3.6 \times 10^{5} M_{\odot}$ (\cite{peterson05}).
In this source, in principle, we could observe the variability due to radiation
pressure instability. As was shown by \cite{czerny09}, the outbursts for the central
black hole of the mass $10^{7} M_{\odot}$, should last below 100 years, so for 
a mass 30 times smaller, the outbursts should last $\sim 3$ years! No such
outbursts are observed. However, this fact is actually consistent with our expectations,
since the Eddington ratio in this source is only $1.2\times 10^{-3}$. The source is
thus stable with both $\alpha P_{\rm tot}$ and $\alpha \sqrt{P_{\rm gas}P_{\rm tot}}$ mechanisms
and provides no useful constraints for the parameterization of the viscous torque.

Significant constraints can be obtained from radio galaxies. In case of the accretion discs in radio galaxies, the Eddington ratios
can be estimated e.g. through the correlation with the broad line luminosities (\cite{dai07}). 
The FR I and FR II sources in this sample
have low Eddington ratios, of 0.00975 and 0.0096, for FR I and FR II sources, respectively.
Observations clearly show that these sources are stable against the radiation pressure 
instability since they form very large scale radio structures. In particular, the central engine of FR II galaxies  must
be operating 
in a continuous way for millions of years. Fig. \ref{fig:topo} shows that 
their stability is consistent with theory if the
heating is given by $\alpha \sqrt{P_{\rm gas}P_{\rm tot}}$.   
On the other hand, the FSRQ sources with compact radio structures tend to have larger Eddington ratios. 
These sources
may in fact exhibit episodic activity and the small size of the structure is
indicating a new episode, as proposed by \cite{czerny09}. Therefore, it seems that the assumption
of the $\alpha \sqrt{P_{\rm gas}P_{\rm tot}}$ can accomodate the observational constraints both for Galactic sources and AGN.

A typical value of the Eddington ratio found in the SDSS sample of quasars 
by \cite{kelly10} 
is 0.05 with a scatter of 0.4 dex. This is also large enough for 
the episodic activity caused by the radiation pressure instability.
Possibly, the selection effect is in fact the reason why we detect only the
sources in the active state: most of 
the sources in the quiescent state are too dim
to be detectable. There is also a possibility that active galaxies at high Eddington ratios,
close to 1, are actually stable due to the stabilizing power of jet/outflow. This mechanism
seems not to work efficiently in Galactic sources but the relative jet power in accreting 
sources rises with the black hole mass:
\begin{equation}
\log L_{R} = 0.6 \log L_{X} + 0.8 \log M
\end{equation}
as was discussed in the context of the so called 
'fundamental plane' of the black hole activity
(\cite{merloni03}, \cite{falcke04}).
Therefore, this effect in AGN can be much stronger than in the Galactic sources.

\section{Discussion}
\label{sec:diss}

The theory of accretion disks suggests the presence of two instabilities:
ionization instability and radiation pressure instability. In the present paper
we made a step towards confronting the theoretical expectations with the observations of Galactic sources and AGN.

The ionization instability is broadly accepted as an explanation of the 
X-ray novae phenomenon and it was compared to the data for galactic s
ources by several authors. In this paper we tested whether the size of 
the disk in the X-ray novae systems is consistent with the conditions 
of the ionization instability. The source SWIFT J1753.3-0127 is at the 
border of the instability strip, which may explain the low outburst 
amplitude in this source. All other sources are located well within the 
instability strip supporting this outburst mechanism.

\begin{figure}
\includegraphics[width=8cm,height=8cm]{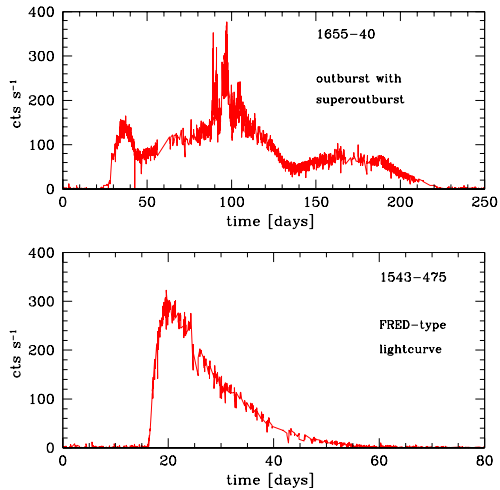}
\caption{RXTE/ASM lightcurves of two X-ray novae. 
The bottom panel shows 4U 1543-475 which is the example of classical FRED behaviour. 
The top panel shows a much more complex behaviour of GRO 1655-40: 
almost symmetric short outburst, followed by extended phase of activity. 
This is possibly an analog of superoutbursts seen in many CV systems}
\label{fig:super}
\end{figure}

The details of the outbursts, however, are not well understood yet.
We note here that the interpretation of the time profiles of X-ray novae
is somewhat complex. Some of the outbursts  have well understood 
 FRED profile (i.e. fast rise and exponential decay), with the sharp luminosity rise 
due to the the ionization instability and an extended wing due to the
X-ray irradiation of the outer disc.
However, in many cases
an additional 'superoutburst' follows the first outburst (see Figure \ref{fig:super}). The possible interpretation
of this secondary, extended maximum may be that the accretion rate  
from the companion star
increases due to some modulation effect (see \cite{smak10} 
for the analysis of
the dwarf novae superoutbursts, very much similar to the X-ray 
novae presented here).
Therefore the classification and quantitative analysis of the outburst durations
due to the ionization instability is not very straightforward.

Interestingly, the famous microquasar GRS1915+105, observed in the
 very high state since
its discovery in 1992, may in fact still be in such a 'superoutburst' phase. 
 On the other hand, the FRED profile is not always seen possibly also due to 
the lack of irradiation of the disc due to unknown reasons. In this case, the ionization instability 
results in a short, symmetric profile. An example of such a source can be 
GRS1730-312, with a timescale of 6 days (see Table \ref{tab:binaries}).

In the case of AGN the applicability of the ionization instability is still under discussion. Fortunately, radio observations can give direct insight into the activity history.
The radio maps of several sources show multiple activity periods, mostly in the form
of double-double structures (\cite{schoenmakers00}, \cite{saripal09}, \cite{marecki09}).
Some of those events may be due to mergers and significant change of the jet direction
suggests such a mechanism. However, if the jet axis does not change, either
a minor merger or ionization instability are the likely cause. The timescales 
can be studied by analyzing the ages of the structures.

Some other sources in turn show a decay phase (\cite{marecki10}). It is 
quite likely that the microquasar, which has been recently discovered to possess
the outflow highly dominated by the kinetic power (\cite{pakull10})
actually also represents such a fading source.
The analysis of such a complex behaviour requires the radio maps with large dynamical range.
Nevertheless, multiple radio surveys are under way and more observational
constraints should be soon available.

The radiation pressure instability model has been
proposed first to model the microquasar time variability
(\cite{taam97}, \cite{nayak00}).
In case of the regular periodic outbursts of
GRS~1915+105 (see e.g. \cite{fb04}), 
lasting from $\sim 100$ to $\sim 2000$ s, 
(depending on the source mean luminosity), 
this approach is successful (\cite{janiuk02}).
No other quantitative mechanism has been put forward to explain the observed 
behavior of this object, and
only the limit cycle mechanism (likely driven by the radiation pressure
instability) explains the absence of the direct transitions from its
 spectral state C to the state B. 

Still, the question arises, why the microquasar seems
to remain an exceptional case where the radiation pressure
instability gives an observational signature.
In the present paper we suggest that several 
other black hole binaries can also be promising objects
for the radiation pressure instability. All the candidate sources have the 
Eddington ratio above $\sim 0.15$. Such a condition is consistent with theory 
if
the viscous torque parameterization as  $\alpha \sqrt{P_{\rm gas}P_{\rm tot}}$ 
is adopted, instead of $\alpha P_{\rm tot}$. Small outburst amplitudes in all
candidate sources (with one exception) also support $\alpha \sqrt{P_{\rm gas}P_{\rm tot}}$  prescription. The recent 3-D MHD simulations show the contribution
to the stress from radiation pressure, but likely weaker than  $\alpha P_{\rm tot}$ (\cite{hirose09b}). Further numerical work and observational constraints 
aimed at finding the proper description of the viscous torque should be treated as complementary.

We argue that the radiation pressure instability also likely applies to active galaxies. The derived separations
between outbursts are on order of $10^6$ yrs for a $10^8\,M_{\odot}$
black hole, while the outburst duration is an order of magnitude
shorter. 

Again the parameterization of the viscosity through 
$\alpha \sqrt{P_{\rm gas}P_{\rm tot}}$ seems to be working better than $\alpha P_{\rm tot}$.  
Such parameterization is consistent with the lack of instability in FR II radio  galaxies as their Eddington ratio is below 0.025 - the lower limit for the instability in active galaxies  if the $\alpha \sqrt{P_{\rm gas}P_{\rm tot}}$ is adopted. This lower limit here  is
much lower than in case of stellar mass objects (see also 
e.g., \cite{sadowski09}) due to the direct dependence on the black hole mass. This model
 has been applied recently also to
explain the apparent young ages of the Giga-Hertz Peak Spectrum
radiosources (\cite{czerny09}). We speculate that in the hot state, the
luminous core will power a radio jet,
while during the cold
state the radio activity ceases.
Scaling the timescale with the black hole mass by a factor $10^8$ gives
the outbursts durations of $10^{2} - 10^{4}$ yrs, and amplitudes are
sensitive to the energy fraction deposited in the jet. 
This gives an 
additional, model-independent argument that the intermittency in quasars on 
the timescales of hundreds/thousands of years is likely of a similar origin
as in the microquasars.    

An important, unstudied aspect is a possible interplay between the two instabilities. The
location of the unstable zone is sensitive to the black hole mass 
(see Fig. \ref{fig:topo}). The two zones are located much closer to each other 
in the case of the active galaxies than in the case of Galactic systems. In 
Galactic binaries only one - ionization instability - 
is frequently operating. 
For small accretion rates, which allow for systematically
lower disc temperatures and the ionization instability fits well in the disc
 size, the radiation pressure instability will not develop.
If both  instabilities are present, they are separated in the disc by over 1000 $R_{S}$ which implies that they are well separated in time, 
acting on timescales of seconds and tends of days, respectively, so they can be modeled independently, as  the radiation pressure instability oscillations of a short 
timescale will be just overimposed on the high luminosity state.

In AGN, the situation is different. 
The partly ionized zone is much closer to the black hole.
If the two unstable zones are very close to each other, the
rate of supply of material to the radiation pressure dominated region may be modulated
on slightly longer timescales, 
independently of the environment changes in the host galaxy.
This poses an observational challenge to deconvolve the interplay between the
two instability timescales, as well as the environmental effects.
Additional modeling effort should thus be  undertaken to
better formulate the theoretical expectations of these instabilities.

Our research has a very preliminary character, and further work is 
clearly needed. On the observational side, careful search for excess 
variability in Galactic sources at timescales of a few tens of seconds 
to a few hundreds of seconds should be done. The main difficulty 
will concern inventing a proper mathematical description of this excess,
because the outbursts due to radiation pressure instability are not likely 
to be strictly periodic and appear clearly in periodograms. 
In the case of the disks around supermassive black holes, more constraints 
should come from detailed radio maps of compact sources showing reactivation 
events in short timescales of hundreds - thousands of years. 
Further development of the models, particularly in the case of the radiation 
pressure instability is also needed, and this should be done in two ways. 
First, if the 3-D MHD computations as those done by Hirose et al. (2009) 
with realistic boundary conditions can be prolonged to viscous timescales, 
we would see whether the instability develops a limit cycle without 
an ad-hoc parametric description of the viscosity. This approach is 
computationally challenging and may not happen soon. Second, the sources 
which in our oppinion are promissing candidates for the radiation pressure 
instability do not exhibit regular outbursts which implies that strong 
non-linear/non-local phenomena are important. Several such phenomena can be 
implemented into the current parametric codes. Irradiation is certainly 
important, and actually some models for ionization instability incorporate 
them. The disk irradiation may also be important for the radiation pressure 
instability. The parameter $\alpha$ may depend locally on the disk thickness, 
and in addition the dissipation may be coupled to the magnetic pressure 
with a radius-dependent significant time delay 
(Kluzniak, private communication).
The theoretical lightcurves can be distorted by a stochastic process, 
such as magnetic dynamo, operating on the local dynamical timescale
(\cite{mayer06}; see also \cite{janiuk07}
for the additional discusssion of this problem in the context of
the magnetically coupled hard X-ray corona). 
That process can evolve e.g. according to the Markov chain model.
The timescale and possibly also the magnetic cell size are governed by the 
$H/R$ ratio. As a result, the outbursts in the limit cycle can be
affected by the flickering, as the fluctuations propagate to the inner disc. 
On the other hand, in between the outbursts the fluctuations are more 
likely to be smeared out. The resulting lightcurve and PDS spectrum 
will depend on the adopted magnitude of the poloidal magnetic field and 
other parameters.
Comparison of such better models with 
observational constraints will help in the future to better understand 
the disk dynamical behaviour.

\section{Conclusions}
\label{sec:concl}

Our preliminary survey of model predictions in confrontation with 
observational data suggests that the parametric description of accretion 
disk viscosity through $\alpha \sqrt{P_{\rm gas}P_{\rm tot}}$ is a promissing 
representation in the radiation pressure dominated disk part. We selected 
several Galactic sources as the candidates which may show the radiation 
pressure instability, but further research is clearly needed. The same law 
likely applies to AGN, and the support comes from the stability of dwarf 
Seyfert galaxy NGC 4395 and FR I and FR II sources. Ionization instability 
criterion in Galactic sources is consistent with the disk sizes. 
There are only very limited constraints for this instability in AGN.

{\bf Acknowledgments} We thank Rob Fender, Ranjeev Misra, 
Marek Abramowicz, Wlodek Kluzniak and Olek Sadowski for helpful discussions.
We are very grateful to the anonymous referee
for his comments which helped us to improve the presentation of the results. 
This work was supported in part by grant NN 203 512638 
from the Polish Ministry of Science. The ASM lightcurves were taken
from http://xte.mit.edu/.

\end{document}